\documentclass[jap, twocolumn, superscriptaddress,showpacs,aps]{revtex4-1}

\usepackage{amsmath}
\usepackage{amstext}
\usepackage{amsfonts}
\usepackage{graphicx}
\usepackage{draftcopy}
\usepackage{flafter}
\usepackage{tabularx}
\usepackage{times}
\usepackage{ mathrsfs }

\def\ket#1{\left|#1\right\rangle}
\def\bra#1{\left\langle#1\right|}
\def\avg#1{\left\langle#1\right\rangle}

\begin{document}
\title{Strong Cavity-Pseudospin Coupling in Monolayer Transition Metal Dichalcogenides: Spontaneous Spin-Oscillations
and Magnetometry}

\author{Amrit De}
\thanks{E-mail: amritde@gmail.com}

\affiliation{Department of Electrical and Computer Engineering, University of California, Riverside, CA 92521}

\author{Roger Lake}

\thanks{E-mail: rlake@ece.ucr.edu}

\affiliation{Department of Electrical and Computer Engineering, University of California, Riverside, CA 92521}

\begin{abstract}
Strong coupling between the electronic states
of monolayer transition metal dichalcogenides (TMDC)
such as MoS$_2$, MoSe$_2$, WS$_2$, or WSe$_2$, and
a two-dimensional (2D) photonic cavity gives rise to several exotic effects.
The Dirac type Hamiltonian for a 2D gapped semiconductor
with large spin-orbit coupling
facilitates pure Jaynes-Cummings type coupling in the presence of a single mode
electric field.
The presence of an additional circularly polarized beam of light gives rise to valley and spin dependent
cavity-QED properties.
The cavity causes the TMDC monolayer to act as an on-chip coherent light source and a spontaneous spin-oscillator.
In addition, a TMDC monolayer in a cavity is a sensitive
magnetic field sensor for an in-plane magnetic field.
\end{abstract}

\date{\today}




\maketitle

Light and matter can become strongly coupled in an optical cavity giving rise to
qualitatively new physics and resulting in numerous applications
in laser physics, optoelectronics, and quantum information processing.
The coherent coupling of light and matter in such systems is described by
cavity quantum electrodynamics (QED).
The advent of quantum information processing has led to significant
activity investigating optical cavity like systems for coherent conversion
of qubits between matter or topological states to phonons, photons and circuit oscillators
\cite{Leibfried2003,Thompson1992,Raimond2001,Hennessy2007,Kovalev2014PRL,Soykal2011,Devoret1989,Schoelkopf2008,Martinis2009}.

Monolayers and bilayers of transition metal dicalcogenides (TMDCs)
couple strongly to light since they are direct bandgap and their large effective masses
result in a large density of states and excitonic binding energies \cite{Jones2013,Xu2014,Jones2014} .
Furthermore, TMDCs
have large spin-orbit (SO) coupling resulting in
spin-valley polarized valence bands \cite{He2014PRL,Chernikov2014PRL}.
The magnetic moment
associated with their valley pseudo-spin gives rise to valley-dependent circular dichroism \cite{Xu2014,Jones2014}.
Access to these valley and spin degrees of freedom can allow for hybrid
on-chip optoelectronic and spintronic devices.
Since the spin degrees of freedom for a band are coupled to a particular valley in momentum space, TMDCs might also
be candidates for qubits with long coherence times, and
there have been suggestions for implementing single qubit gates in TMDC quantum dots \cite{Kormanyos2014}
and bilayers \cite{Gong2013}.
%

\begin{figure}[t]
  \centering
\includegraphics[width=1\columnwidth]{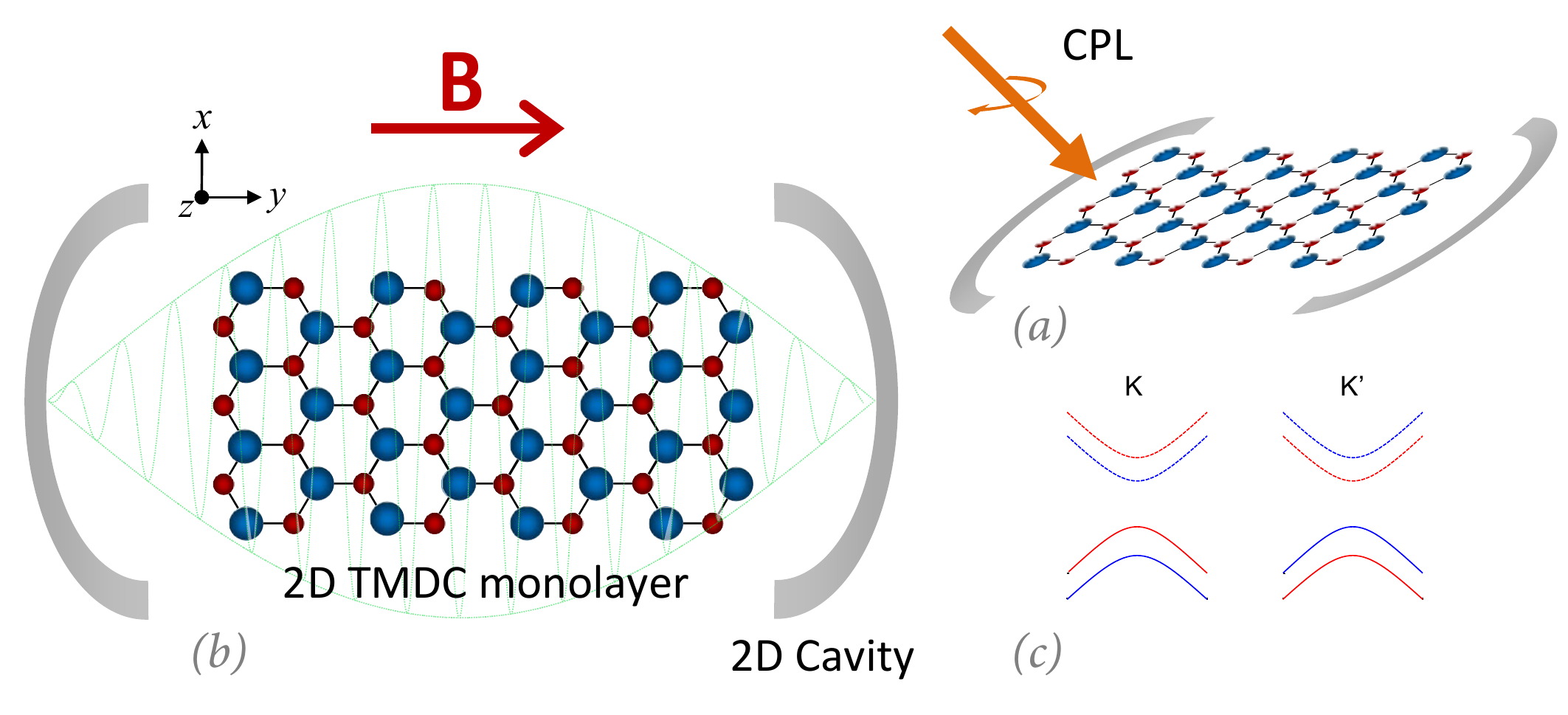}
\caption{Schematic of a monolayer TMDC in a 2D photonic cavity (a) with an incident beam of circularly polarized
light(CPL) to break the valley degeneracy and (b) with an in-plane magnetic field for magnetometry.
(c) Sample bandstructure about $K$ and $K'$ showing the breaking of the conduction band degeneracy from the
effective inplane magnetic field. Colors indicate the band's spin.  }
\label{fig:scheme}
\end{figure}

Recent experiments have demonstrated strong coupling effects between 2D materials and photonic cavities.
Graphene (Gr) can couple to a photonic crystal's evanescent modes \cite{Gan2012}.
It has also been
suggested that quantum two level systems (TLSs) can be coupled to surface plasmon modes in
graphene \cite{Koppens2011}.
A Gr/TMDC/Gr heterostructure can be used for photovoltaics
\cite{Britnell2013}, and an excitonic laser can be created by placing monolayer MoS$_2$ in a cavity
\cite{Ye2015Nature}.

In this paper, we show that new properties and device functionalities arise
when strong light-matter interaction can be achieved for a
gapped 2D Dirac material with strong SO splitting in a 2D optical cavity.
2D cavities are now quite common in circuit-QED \cite{Schoelkopf2008}.
Usually in cavity-QED, the TLS-field coupling occurs via a dipole term or
through some nonlinear interactions.
Here, the linear k-dependent 2D Dirac type Hamiltonian, after a canonical
transformation, directly gives Rabi- or Jaynes-Cummings type coupling
between the cavity field and the lattice pseudo-spin.
In TMDCs, since the valley and spin indices are coupled, each valley can be
selected using circularly polarized light (CPL) of a given handedness \cite{Xiao2012PRL}.
In a suitable optical cavity this
could lead to spontaneous spin oscillations for spintronics.
However,  CPL flips its handedness upon
reflection from a conventional surface -- hence it is not easy
to design a cavity that will only sustain ether just left- or
right-CPL.
Some reflective chiral surfaces can suppress this cross polarization \cite{Hodgkinson2002,Maksimov2014}.
The evanescent mode of a chiral photonic crystal \cite{Konishi2011} could also be used,
but then the TLS-cavity coupling
will not be strong.

We suggest a simpler approach and point out an additional advantage of the 2D architecture.
In the strongly coupled system of
a 2D TMDC monolayer inside a 2D cavity,
spontaneous vacuum Rabi oscillations occur even in
the absence of photons.
However, a 2D cavity only supports linearly polarized electromagnetic vibrational modes.
Hence the oscillations occur for both valleys with opposite spin.
We propose introducing an additional CPL beam that is incident on the 2D cavity-QED system.
Sufficiently intense CPL will select a valley.
Since intravalley transitions conserve spin,
this leads to pure spin Rabi flopping.
In the absence of cavity photons, the spin Rabi oscillations
are spontaneous in the strong coupling regime.
The vacuum Rabi oscillations for the the spin
polarization (SP) of a given valley are amplitude modulated.
The overall degree
of valley polarization depends on the CPL's intensity which blue-shifts the Rabi frequency.
At higher photon numbers, for a field in a coherent state, the
expected collapse and revival type behavior for the Rabi oscillations can be seen.
In the absence of a cavity mode that
supports CPL, say in a strictly 2D geometry, this device can be used as an on-chip coherent light source and a
frequency comb at higher photon number.

This cavity-pseudospin system also has other applications.
 The system is a highly sensitive sensor of an in-plane magnetic field.
An in-plane magnetic field shifts the vacuum Rabi frequency.
An extremely encouraging finding is that this frequency shift is scale invariant.
Whereas one usually seeks strong cavity-TLS coupling,
here the scale invariance suggests high sensitivity even
for weak coupling.

\begin{center}
\begin{table}
\renewcommand{\arraystretch}{1.5}
\begin{tabular}{c|c c}
 \hline
 \hline
{Material} & $ \Lambda_-$ (GHz) & $\Lambda_+$ (GHz)  \\
\hline
MoS$_2$ & 6.411 &  5.898  \\
WS$_2$  &  14.021  &  10.366 \\
MoSe$_2$ &  6.128  &  5.442 \\
WSe$_2$  &  8.29  &  6.295 \\
\hline
\end{tabular}
\caption {Coupling strength or the vacuum Rabi frequency for various TMDCs at zone center. Here $\Lambda_\pm$
corresponds to $\omega_\pm=E_g\pm\Delta_{so}$ transition frequencies. The 2D cavity mode volume($V$) is assumed
to be $20\mu m\times 20\mu m\times 100 nm$. Note that $\Lambda_\pm\propto V^{-1/2}$.}
\label{tab:lam}
\end{table}
\end{center}

\noindent
\emph{The Model}:
Consider the Hamiltonian for a monolayer TMDC in an in-plane magnetic field along $y$.
 \begin{eqnarray}
 H_o' =  H_o + H_b
 \end{eqnarray}
$H_o$ is the following effective $4\times4$ two-band $k \cdot p$ Hamiltonian for a given valley,
\begin{eqnarray}
H_{o} = u(\tau\tilde\sigma_xk_x +\tilde\sigma_yk_y) + \frac{E_g}{2}\tilde\sigma_z + \frac{\Delta_{so}}{2}\tau s_z(\tilde
\sigma_z-I)~~~~
\label{Ho}
\end{eqnarray}
where $E_g$ is the band gap,
$\Delta_{so}$ is the SO splitting and $u$ is the velocity.
Here $H_b=g\mu_B B_y s_y$ where $g$ is the g-factor and $\mu_B$ is the Bhor-magneton.
We use $g=-4$ \cite{Wang20152D}.
The Pauli spin matrices along $j$ are $s_j$,
and
$\tilde\sigma_j$ are the pseudo-spin Pauli matrices in the orbital basis,
$\{\psi_c,\psi_v^\tau\} = \{\ket{d_{z^2}}, \ket{d_{x^2+y^2}+i\tau d_{xy}}\}$.
It is implied that $\tilde\sigma_j=I\otimes
\sigma_j$ and $s_j=\sigma_j\otimes I$.

First consider the case of a monolayer-TMDC in a cavity with $B_y=0$.
For a reflective cavity, with the single-mode of an
electric field oscillating along $\hat x$, one can canonically transform $k_{x}\rightarrow k_{x}+A_{x}$, where $A_{x}$ is
the vector potential along $x$.
The TDMC's coupling Hamiltonian will be $H_{i} = u\tau\sigma_x A_x(t)= \Lambda\tau
\sigma_x(a^*e^{i\omega t} + ae^{-i\omega t})$ where $\Lambda$ is the coupling constant between the cavity's electric
field and the TMDC bands.
Second quantizing the cavity field and making the rotating wave approximation,
$\Lambda \sigma_x(a^{\dagger} + a ) \approx\Lambda(a^{\dagger}\sigma_- + a\sigma_+)$,
where $a^\dagger(a)$ are the photon
creation(annihilation) operators,
we obtain a block diagonal Hamiltonian in the dressed orbital state basis.
Hence the
total system Hamiltonian is
\begin{eqnarray}
H= H_o +\frac{\omega}{2}{a^{\dagger}a} + \Lambda\tau(a^{\dagger}\sigma_- + a\sigma_+) .
\end{eqnarray}
The complete basis set now consists of the spinor part and the dressed orbital states,
$\{[\ket{+},\ket{-}] \otimes[\ket{\psi_{c},n},\ket{\psi_{v}^{\tau},n+1}]\}$.
Note that cavity part and the spin part($\ket{+},\ket{-}$) are decoupled in the absence of an external in-plane magnetic
field.

We estimate the field coupling constant $\Lambda$ assuming a simple 2D rectangular cavity.
The coupling constant for four different TMDCs on resonance are calculated as $\Lambda\propto u\sin(\kappa y)
[\epsilon_oV(E_g\pm\Delta_{so})]^{-\frac{1}{2}}$,
where $\kappa$ is the electromagnetic wavevector and $V$ is the mode volume.
If $V$ is small, then the coupling will be strong.
For the mode volume we assume a nearly-2D rectangular cavity.
Hence one can anticipate that strong
coupling can be achieved between a 2D TMDC and the cavity.
The $\Lambda$s for the two different transitions just for a
monolayer TMDC are shown in table.-\ref{tab:lam}.

\begin{figure}
\includegraphics[width=1\columnwidth]{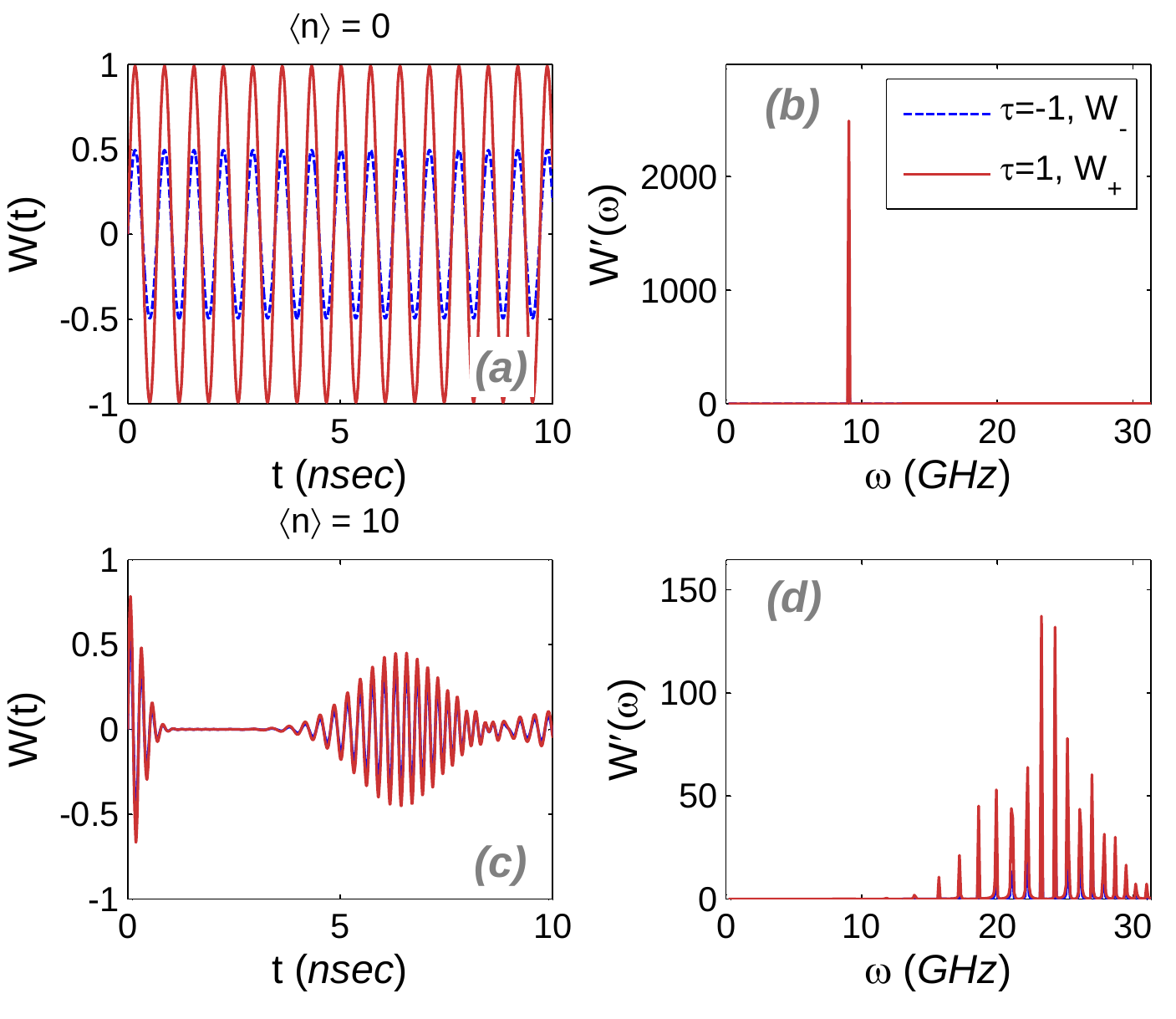}
\caption{Population inversion for the $\tau=\pm$ valleys showing (a) vacuum Rabi oscillations for $\avg{n}=0$ and (c)
collapse and revival of Rabi oscillations for $\avg{n}=10$ for a field initially in a coherent state. The corresponding power
spectra, $\mathcal{W}'(\omega)=\int|{\mathcal W}(t)|^2e^{i\omega t}dt$, is shown in (b) and (d). Here the drive is $
\omega=E_g-\Delta_{so}$, and the circularly polarized light has $\avg{{E}}=\Lambda$.}
\label{fig:Rabi_Ak0}
\end{figure}

The direct product of the photon state
and the valance band wavefunction at initial time $t=0$ is
$\Psi_{\pm,{\bf k}}^{\tau,v} = \sum C_n |\psi_v^\tau;\pm ; n\rangle$.
where $|C_n|^2$ is the probability distribution number of $n$ photons.
The wavefunction at time $t$ is obtained by time evolving with $U = \exp(-iHt)$.

Since the two valleys do not couple with each other one can only consider intravalley optical effects in the present
model.
For a given valley, in the absence of an external magnetic field, the valance to conduction band(CB) {population
inversion} in the cavity is
\begin{equation}
\mathcal{W^{\tau}_\pm}({\bf k}) =  \displaystyle\sum_n \left|C_n\right|^2\left(\vartheta_\pm\cos(\frac{\nu^{\tau}_\pm t}
{2}) + \varphi_\pm\sin(\frac{\nu^{\tau}_\pm t}{2})\right),
\end{equation}
where
%
\begin{align}
  \vartheta_\pm &= \frac{1 + (\mathcal{E_\pm}^2-\mathcal{L}_\pm^2)} { [(1+(\mathcal{E_\pm}+\mathcal{L}_\pm)^2)(1+
(\mathcal{E_\pm}-\mathcal{L}_\pm)^2)]^{1/2}  }
\nonumber \\
\varphi_\pm &= \frac{w_\pm(\mathcal{E_\pm}+\mathcal{L}_\pm) + w_\pm^*(\mathcal{E_\pm}-\mathcal{L}_\pm)  +f_
\pm(\mathcal{E_\pm}^2-\mathcal{L}_\pm^2-1)} { [(1+(\mathcal{E_\pm}+\mathcal{L}_\pm)^2)(1+(\mathcal{E_\pm}-
\mathcal{L}_\pm)^2)]^{1/2}  }
\nonumber \\
\mathcal{E_{\pm}}  &= {(E_g \pm {\tau}\Delta_{so})}/{2({\tau k_x'+ik_y'})}
\nonumber  \\
\mathcal{L_{\pm}}  &= {\sqrt{\left({E_g}\pm{{\tau}\Delta_{so}}\right)^2 + 4(k_x'^2+k_y'^2)}}/{2(\tau k_x'+ik_y')}.
\nonumber
\end{align}
%
Here $\Lambda_n = \Lambda\sqrt{1+n}$, $k'_{x(y)}=u k_{x(y)}$,
$\Omega^{\tau}_\pm = E_g\pm{\tau}\Delta_{so} -\omega$,
$\nu^{\tau}_\pm = \sqrt{(\Omega^{\tau}_\pm)^2 +(\tau k'_x + {\tau}\Lambda_n)^2 + k_y'^2}$,
$f_\pm = {\Omega^{\tau}_\pm}/{\nu^{\tau}_\pm}$ and
$w_\pm = {(\tau\Lambda_n + \tau k_x' -ik_y' )}/{\nu^{\tau}_\pm}$.
The $\pm$ signs represent different spin states.
Since these transitions are also $\bf k$ dependent, the overall inversion probability is obtained after integrating over $\bf
k$, $\mathcal{W}^{\tau}_\pm = \displaystyle\int\mathcal{W^{\tau}_\pm}({\bf k})d{\bf k}$.

\begin{figure}
\includegraphics[width=1\columnwidth]{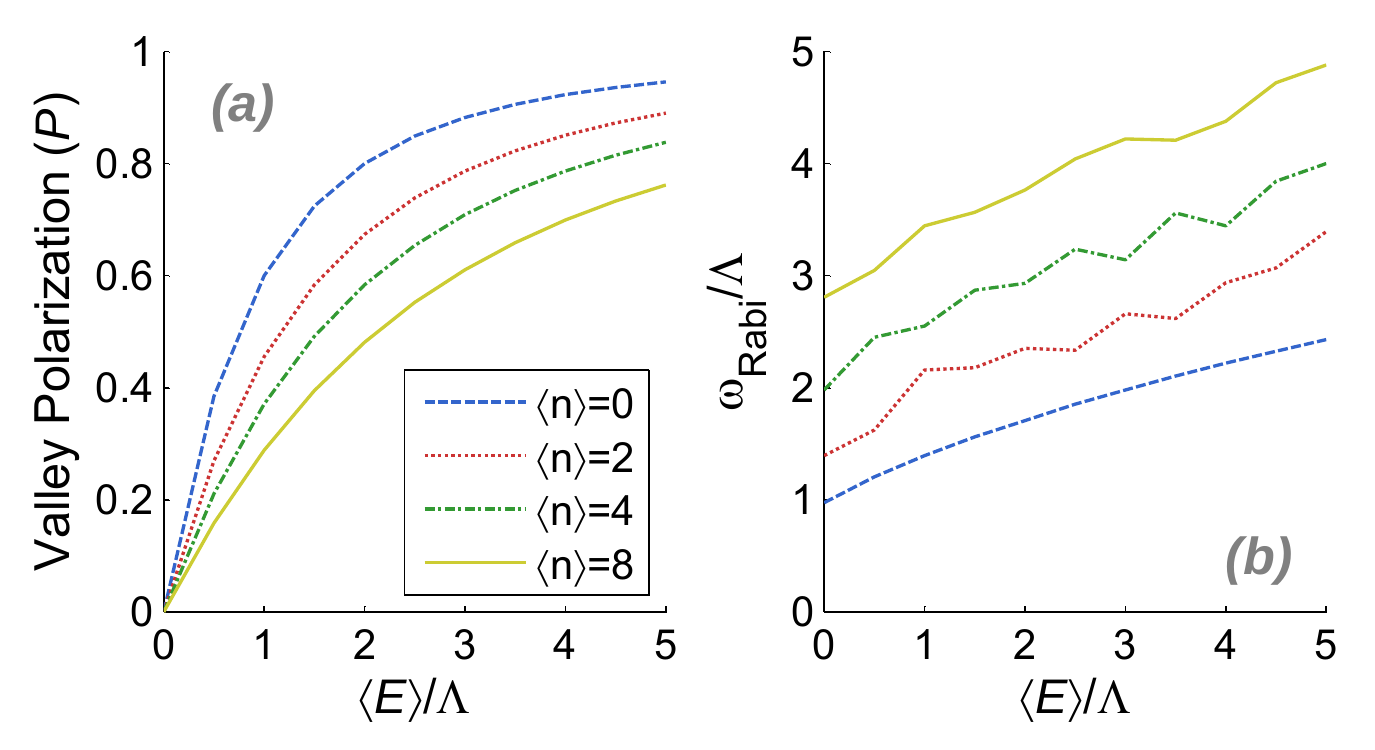}
\caption{(a) Valley polarization inside a cavity as a function of CPL intensity and for different photon numbers. (b) Peak
Rabi frequency shift as a function of CPL intensity. The cavity's resonance frequency is $\omega=E_g-\Delta_{so}$.}
\label{fig:VP}
\end{figure}

\emph{Valley Selectable Photonic Spin Oscillator:}
In the absence of $B_y$, in a monolayer TMDC, only interband transitions between bands of the same spin are allowed.
We assume that the single cavity mode is initially in a coherent state, $|C_n|^2=\exp(-\langle n \rangle)\frac{{\langle n
\rangle}^{n}}{n!}$
where $\langle{n}\rangle$ is the average photon number.

For a drive resonant with the gap, $\omega=E_g-\Delta_{so}$,
the vacuum Rabi oscillations are shown in Fig. \ref{fig:Rabi_Ak0}(a)
along with the corresponding Fourier spectra in Fig. \ref{fig:Rabi_Ak0}(b).
For $\langle{n}\rangle=0$ there is spontaneous emission and Rabi flipping for the TLS.
For the opposite spin states (with gap $E_g+\Delta_{so}$),
the maximum Rabi oscillation amplitude $\sim 0$ since $\omega$ is off-resonance with
this transition.

The valley dependent Rabi oscillations are shown in Fig. \ref{fig:Rabi_Ak0}.
These are pure spin oscillations from the cavity coupling.
A bias for a particular valley (hence spin) is created by using CPL of a given handedness.
We use an additional canonical transformation for introducing CPL,
$k'_x \rightarrow k'_x + \avg{{E}_x} $
and $ k'_y \rightarrow k'_y \pm i\avg{{E}_y}$ where, $\avg{{E}_x}= \avg{{E}_y}=\avg{{E}}$ and $\avg{{E}}$ is the time
averaged field.
Right-CPL favors the $\tau=1$ valley as shown in Fig. \ref{fig:Rabi_Ak0} since it biases the Hamiltonian by adding a $
\avg{{E}_x}\tau\tilde\sigma_{x} + i\avg{{E}_y}\tilde\sigma_{y}$ term.

When $\langle{n}\rangle>0$,
the Rabi oscillations undergo collapse and revival (CR) which are more rapid, more
distinct, and temporally spaced further apart with increasing $\avg{n}$.
Each term in the summation over $n$ represents
Rabi flips weighted by $C_n$, which are all correlated at $t=0$.
However, at longer times the destructive interference
between the weighted terms leads to the collapses and then constructive interference leads to revivals.
Fig. \ref{fig:Rabi_Ak0}(c) shows that this purely quantum mechanical CR feature can be individually observed for each
valley.
This CR happens even in the presence of CPL,
but the amplitude of the CRs are inequivalent for each valley, and
they continue indefinitely with each revival being smaller in amplitude and less distinct from the preceding
collapse.
The Fourier spectra shows a \emph{frequency comb} type behavior, where the number of spectral peaks is $
\propto \avg{n}$. In the present treatment CPL blue-shifts the central Rabi frequency peak, which is discussed in greater
detail next.

The valley(spin) polarization in this system can be characterized as follows,
\begin{equation}
\mathcal{P}=  \frac{|\mathcal{W}^{\tau=1}_+|^2_{max} - |\mathcal{W}^{\tau=-1}_-|^2_{max}}{|\mathcal{W}^{\tau=1}_+|
^2_{max} + |\mathcal{W}^{\tau=-1}_-|^2_{max}} .
   \label{eq.VP}
\end{equation}
The degree of valley polarization depends on the CPL's intensity $\avg{\mathcal E}$
as shown in Fig. \ref{fig:VP}(a).
As $\avg{\mathcal E}$ is increased $\mathcal{P}$ tends towards 1, but it also tends to saturate.
For a given $\avg{\mathcal E}$, $\mathcal{P}$ is higher for smaller $\avg{n}$.
This is because the cavity photons are also vibrating along $x$
which will tend to make the incident light more elliptically polarized as $\avg{n}$ is increased.
Increasing $\avg{\mathcal{E}}$ also shifts the central Rabi frequency peak towards
higher frequencies as shown in Fig. \ref{fig:VP}(b),
since CPL increases the effective $\Lambda$ (see $\nu_\pm^\tau$).
The slope $d\mathcal{P}/d\avg{\mathcal{E}}$ is
roughly the same for all $\avg{n}$ as expected.

A spin-Zeeman field along $z$ does not affect any of these results,
since it simply adds a phase factor.
For fields up to $1$ T the valley Zeeman effect \cite{Aivazian2015Nat} also does
not do much although it polarizes the valleys.
The addition of an inplane magnetic field $B_y$ however leads to level detuning and leakages
-- but again it does not affect
the overall $\mathcal{P}$ for this system for fields up to $1$ T.
However the inclusion of $B_y$ in this 2D cavity-QED system reveals a key technological application.

\begin{figure}
\includegraphics[width=1\columnwidth]{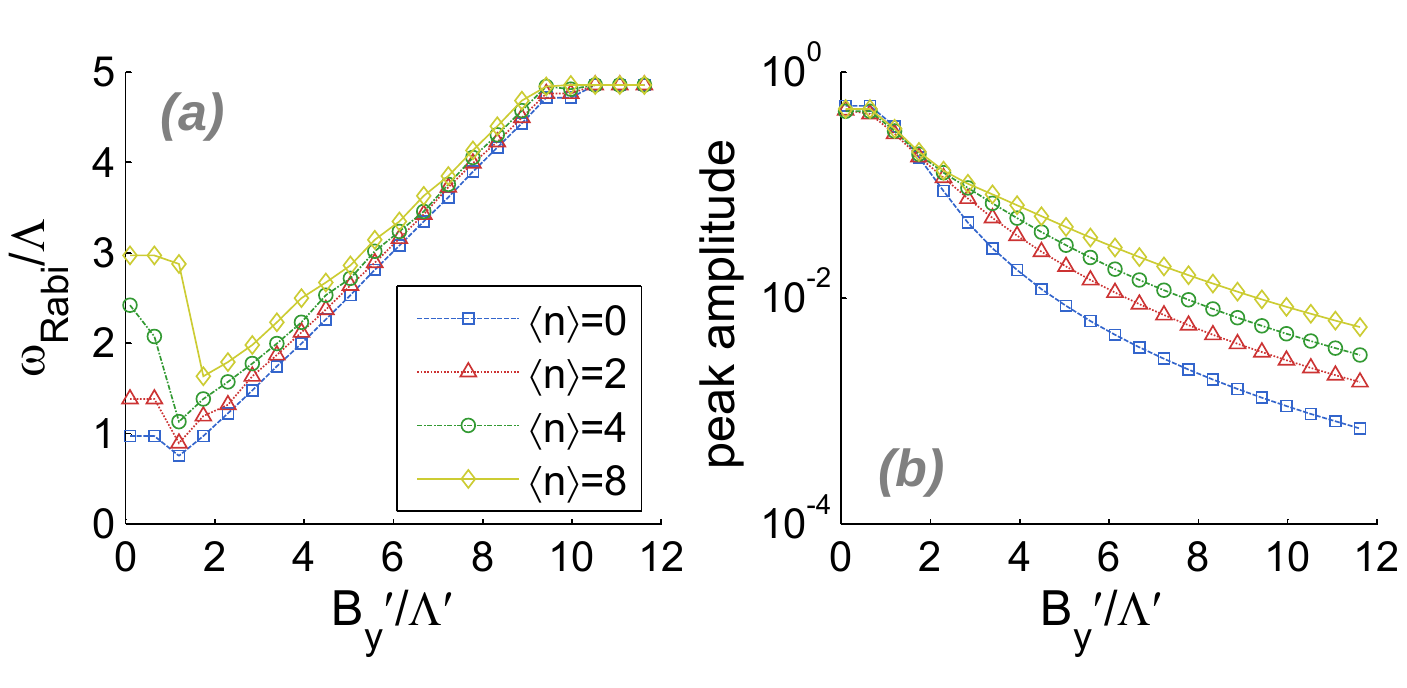}
\caption{ (a) Normalized peak Rabi frequency as a function of the normalized magnetic field for different photon
numbers. (b) Corresponding peak oscillation amplitude for the population transfer between valance- and conduction
band states with the same spin for the $\tau=1$ valley. The drive $\omega=E_g-\Delta_{so}$, $\Lambda^\prime=\hbar
\Lambda$ and $B^\prime_y=g\mu_BB_y$.}
\label{fig:Peak2}
\end{figure}

\emph{Application as a Magnetometer:}
A monolayer TMDC in an optical cavity can be used for sensitive magnetic field sensing applications.
For a 2D material, $B_y$ only affects the spin states and does not lead to the formation of Landau levels.
However, now the two orbital TLSs are not deoupled anymore
and the dynamics of this system is significantly altered in the presence of $B_y$.
We numerically calculate the
population transfer probabilities in the $4\times4$ the dressed state basis.

The magnetic field modifies the zone center energies to $E_g/2\pm B_y$ and
$- E_g/2 \mp \sqrt{B_y^2 + \Delta_{so} ^2}$,
which leads to level detuning.
This leads to an increase in the Rabi oscillation frequency and a decrease in the
oscillation amplitude.
The peak Rabi-flop frequencies normalized to $\Lambda$ are shown as a function of $B_y/\Lambda$
for different photon numbers for the drive $\omega=E_g-\Delta_{so}$ in Fig. \ref{fig:Peak2}.

At low $B_y$, $\omega_{peak}$ does not change much.
And at higher $B_y$, $\omega_{peak}$ saturates.
However the results in the intermediate regime are extremely encouraging.
First, the linear scaling of $\omega_{peak}$ with $B_y$,
and the invariance of this linear scaling and its slope with
$B_y/\Lambda$ implies that this device can be used
as a very sensitive magneto-meter.

This is a key result.
For most cavity-QED applications such as lasing, very strong coupling is desired.
Here because of
the invariance as a function of $B_y / \Lambda$ one could get to very small
magnetic field sensing limits.
This is only possible because of the unique combination of a
gapped material with large SO interactions in a 2D geometry -- all of
which are necessary.
The direct gap makes the system optically active and the 2D geometry allows $B_y$ to couple to
spin without introducing unwanted Landau levels which then subsequently couples the CB orbitals.

In theory these effects can be reproduced if one just added a $B_y\sigma_y$ term
to the Jaynes-Cummings Hamiltonian.
But physically one cannot have an electric dipole coupling and a
magnetic field coupling in the same matrix element for an orbital two-level-system.
We also argue that this would be robust even with cavity imperfections and
spin-dephasing and relaxation as one is not concerned with the decay of the signal,
but just with the main peak Fourier component.
Experimentally this would amount to spectrally decomposing the time dependent photo-luminescence
signal.

\begin{figure}
\includegraphics[width=1\columnwidth]{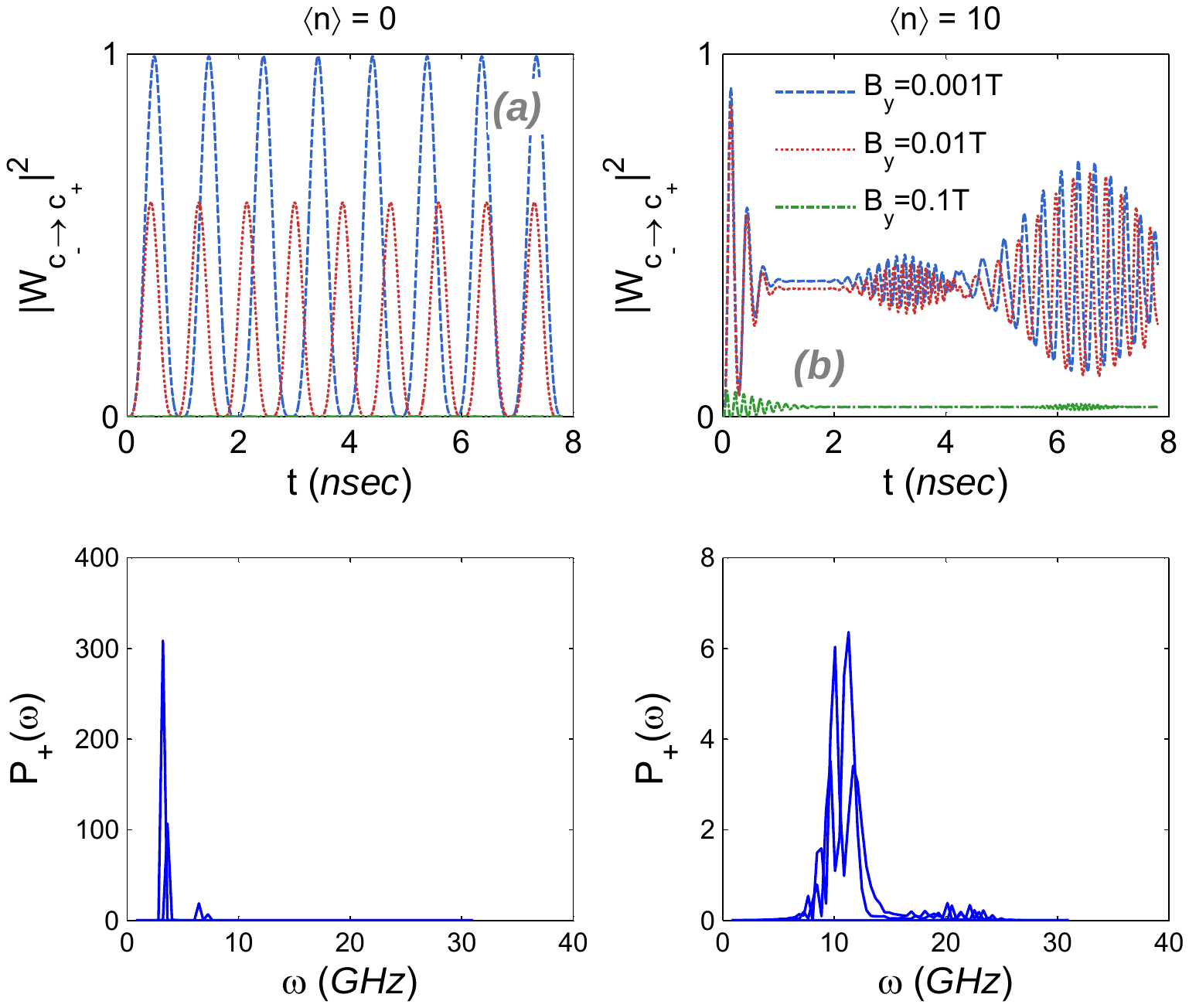}
\caption{Population transfer between the conduction band states with opposite spin $\psi_{c,-}\leftrightarrow \psi_{c,+}$
shown for two different photon numbers and magnetic fields. The drive is at $\omega=E_g-\Delta_{so}$.}
\label{fig:vmvp}
\end{figure}

\emph{Unusual Conduction Band Transitions:}
In the presence of $B_y$, either direct or indirect transitions between
all four states in a valley are allowed.
However some rather peculiar features stand out for the CB
$\psi_{c,-}\leftrightarrow \psi_{c,+}$ transitions.
The Rabi flops for these transitions are shown in Fig. \ref{fig:vmvp}.
These results were obtained exactly by numerically projecting
$U=\exp(-iHt)$ between the dressed CB eigenstates of $H_o'$.
At very small but finite $B_y$, the vacuum Rabi flops reach 1.
As the magnetic field strength increases, the amplitude decreases, but the
Rabi frequency does not shift.
But again if $B_y=0$, this Rabi flipping would vanish.

In general $[H_0, H_b] \neq 0$,
however to gain a more intuition one can approximate
$U \approx e^{-iHt}e^{-iH_b t}$,
which is valid for small $B_y$.
Then the CB population inversion is
\begin{align}
\bra{\Psi_-^{c}}U\ket{\Psi_+^{c}} & \sim  \cos(B_y t)(\mathcal{W'}_+ - \mathcal{W'}_-)
\nonumber \\
 & + i\sin(B_y t)(\mathcal{C}_+\mathcal{W'}_+ - \mathcal{C}_-\mathcal{W'}_-)
\label{eq:WB}
\end{align}
where $\mathcal{C}_{\pm} = (\sqrt{B_y^2+\Delta_{so}^2} \pm \tau\Delta_{so})/B_y$
and $\mathcal{W'}_\pm = \sum |C_n|^2\left(\cos(\frac{\nu_\pm t}{2}) +\frac{\Omega_\pm}{\nu_\pm}\sin(\frac{\nu_\pm t}
{2})\right)$.
Eq. \ref{eq:WB} approaches 1 in the limit of a vanishing $B_y$.
The peculiar Rabi flopping in Fig. \ref{fig:vmvp} can be explained as follows.
In the absence of a magnetic field,
the CB $c_\pm$ states are degenerate but are
completely decoupled from each other in the present $k \cdot p$ model.
An infinitesimally small $B_y$ lifts this
degeneracy and couples the two CB states allowing
$\Psi^c_{-}\leftrightarrow \Psi^c_{+}$ transitions.
However now since the two levels are still nearly degenerate for a small $B_y$,
there is an almost perfect overlap of the
wavefunctions.
As a result the Rabi flops reach 1 for infinitesimally small $B_y$s.

Eq. \ref{eq:WB} reproduces this behavior in the limit of small $B_y$.
As the magnetic field strength increases,
the amplitude decreases, but the Rabi frequency does not shift.
For $\avg{n}=10$, the CR type behavior is seen for these states also.

It should be noted that in actual TMDC monolayers, the CB states are also non-degenerate and
spin-split, although the spin splitting is much less than that of the valence
band \cite{Heine_AnnPhys14,Kormanyos2014,Kormanyos2015}.
%
%
%
While this CB splitting does not affect the other results, the CB population transfer amplitudes would less than what is shown in Fig. \ref{fig:vmvp}.
Overall, this is an important effect to take into account when considering quantum information processing applications.

In summary we show that the strong coupling between the lattice pseudo-spin of an inversion
asymmetric monolayer TMDC and a photonic cavity leads to a number of physical effects.
This system can act as an on-chip coherent light source.
The Dirac type Hamiltonian for a 2D gapped
semiconductor with a large SO interaction facilitates pure Jaynes-Cummings
coupling in the presence of a single mode electric field.
This gives rise to valley and spin dependent optical properties which can
be controlled in a 2D architecture by using an additional CPL field.
With CPL, the strong coupling effect leads to a \emph{spontaneous} spin-oscillator with
zero cavity photons!
For a higher photon number coherent state field, valley selective collapse and revival of Rabi
oscillations occur.

The presence of an external in-plane magnetic field leads to a number of additional effects.
This TMDC cavity-QED device can be used for sensitive magnetic field sensing applications,
which is only possible here because of the
combination of a gapped 2D material with large SO interactions in a cavity.
A consequence of an inplane magnetic field
is that Rabi oscillations between opposite CB spin states
become feasible in a monolayer-TMDC.
These oscillations are robust for
small magnetic field fluctuations which could also be
useful for quantum information applications.

\noindent
{\em Acknowledgements:}
This work was supported by FAME, one of six centers of STARnet,
a Semiconductor Research Corporation program sponsored by MARCO and DARPA
and the NSF 2-DARE, EFRI-143395.


\begin{thebibliography}{28}%
\makeatletter
\providecommand \@ifxundefined [1]{%
 \@ifx{#1\undefined}
}%
\providecommand \@ifnum [1]{%
 \ifnum #1\expandafter \@firstoftwo
 \else \expandafter \@secondoftwo
 \fi
}%
\providecommand \@ifx [1]{%
 \ifx #1\expandafter \@firstoftwo
 \else \expandafter \@secondoftwo
 \fi
}%
\providecommand \natexlab [1]{#1}%
\providecommand \enquote  [1]{``#1''}%
\providecommand \bibnamefont  [1]{#1}%
\providecommand \bibfnamefont [1]{#1}%
\providecommand \citenamefont [1]{#1}%
\providecommand \href@noop [0]{\@secondoftwo}%
\providecommand \href [0]{\begingroup \@sanitize@url \@href}%
\providecommand \@href[1]{\@@startlink{#1}\@@href}%
\providecommand \@@href[1]{\endgroup#1\@@endlink}%
\providecommand \@sanitize@url [0]{\catcode `\\12\catcode `\$12\catcode
  `\&12\catcode `\#12\catcode `\^12\catcode `\_12\catcode `\%12\relax}%
\providecommand \@@startlink[1]{}%
\providecommand \@@endlink[0]{}%
\providecommand \url  [0]{\begingroup\@sanitize@url \@url }%
\providecommand \@url [1]{\endgroup\@href {#1}{\urlprefix }}%
\providecommand \urlprefix  [0]{URL }%
\providecommand \Eprint [0]{\href }%
\providecommand \doibase [0]{http://dx.doi.org/}%
\providecommand \selectlanguage [0]{\@gobble}%
\providecommand \bibinfo  [0]{\@secondoftwo}%
\providecommand \bibfield  [0]{\@secondoftwo}%
\providecommand \translation [1]{[#1]}%
\providecommand \BibitemOpen [0]{}%
\providecommand \bibitemStop [0]{}%
\providecommand \bibitemNoStop [0]{.\EOS\space}%
\providecommand \EOS [0]{\spacefactor3000\relax}%
\providecommand \BibitemShut  [1]{\csname bibitem#1\endcsname}%
\let\auto@bib@innerbib\@empty
\bibitem [{\citenamefont {Leibfried}\ \emph {et~al.}(2003)\citenamefont
  {Leibfried}, \citenamefont {Blatt}, \citenamefont {Monroe},\ and\
  \citenamefont {Wineland}}]{Leibfried2003}%
  \BibitemOpen
  \bibfield  {author} {\bibinfo {author} {\bibfnamefont {D.}~\bibnamefont
  {Leibfried}}, \bibinfo {author} {\bibfnamefont {R.}~\bibnamefont {Blatt}},
  \bibinfo {author} {\bibfnamefont {C.}~\bibnamefont {Monroe}}, \ and\ \bibinfo
  {author} {\bibfnamefont {D.}~\bibnamefont {Wineland}},\ }\href {\doibase
  10.1103/RevModPhys.75.281} {\bibfield  {journal} {\bibinfo  {journal} {Rev.
  Mod. Phys.}\ }\textbf {\bibinfo {volume} {75}},\ \bibinfo {pages} {281}
  (\bibinfo {year} {2003})}\BibitemShut {NoStop}%
\bibitem [{\citenamefont {Thompson}\ \emph {et~al.}(1992)\citenamefont
  {Thompson}, \citenamefont {Rempe},\ and\ \citenamefont
  {Kimble}}]{Thompson1992}%
  \BibitemOpen
  \bibfield  {author} {\bibinfo {author} {\bibfnamefont {R.~J.}\ \bibnamefont
  {Thompson}}, \bibinfo {author} {\bibfnamefont {G.}~\bibnamefont {Rempe}}, \
  and\ \bibinfo {author} {\bibfnamefont {H.~J.}\ \bibnamefont {Kimble}},\
  }\href {\doibase 10.1103/PhysRevLett.68.1132} {\bibfield  {journal} {\bibinfo
   {journal} {Phys. Rev. Lett.}\ }\textbf {\bibinfo {volume} {68}},\ \bibinfo
  {pages} {1132} (\bibinfo {year} {1992})}\BibitemShut {NoStop}%
\bibitem [{\citenamefont {Raimond}\ \emph {et~al.}(2001)\citenamefont
  {Raimond}, \citenamefont {Brune},\ and\ \citenamefont
  {Haroche}}]{Raimond2001}%
  \BibitemOpen
  \bibfield  {author} {\bibinfo {author} {\bibfnamefont {J.~M.}\ \bibnamefont
  {Raimond}}, \bibinfo {author} {\bibfnamefont {M.}~\bibnamefont {Brune}}, \
  and\ \bibinfo {author} {\bibfnamefont {S.}~\bibnamefont {Haroche}},\ }\href
  {\doibase 10.1103/RevModPhys.73.565} {\bibfield  {journal} {\bibinfo
  {journal} {Rev. Mod. Phys.}\ }\textbf {\bibinfo {volume} {73}},\ \bibinfo
  {pages} {565} (\bibinfo {year} {2001})}\BibitemShut {NoStop}%
\bibitem [{\citenamefont {Hennessy}\ \emph {et~al.}(2007)\citenamefont
  {Hennessy}, \citenamefont {Badolato}, \citenamefont {Winger}, \citenamefont
  {Gerace}, \citenamefont {Atature}, \citenamefont {Gulde}, \citenamefont
  {Falt}, \citenamefont {Hu},\ and\ \citenamefont {Imamoglu}}]{Hennessy2007}%
  \BibitemOpen
  \bibfield  {author} {\bibinfo {author} {\bibfnamefont {K.}~\bibnamefont
  {Hennessy}}, \bibinfo {author} {\bibfnamefont {A.}~\bibnamefont {Badolato}},
  \bibinfo {author} {\bibfnamefont {M.}~\bibnamefont {Winger}}, \bibinfo
  {author} {\bibfnamefont {D.}~\bibnamefont {Gerace}}, \bibinfo {author}
  {\bibfnamefont {M.}~\bibnamefont {Atature}}, \bibinfo {author} {\bibfnamefont
  {S.}~\bibnamefont {Gulde}}, \bibinfo {author} {\bibfnamefont
  {S.}~\bibnamefont {Falt}}, \bibinfo {author} {\bibfnamefont {E.~L.}\
  \bibnamefont {Hu}}, \ and\ \bibinfo {author} {\bibfnamefont {A.}~\bibnamefont
  {Imamoglu}},\ }\href {\doibase 10.1038/nature05586} {\bibfield  {journal}
  {\bibinfo  {journal} {Nature}\ }\textbf {\bibinfo {volume} {445}},\ \bibinfo
  {pages} {896} (\bibinfo {year} {2007})}\BibitemShut {NoStop}%
\bibitem [{\citenamefont {Kovalev}\ \emph {et~al.}(2014)\citenamefont
  {Kovalev}, \citenamefont {De},\ and\ \citenamefont
  {Shtengel}}]{Kovalev2014PRL}%
  \BibitemOpen
  \bibfield  {author} {\bibinfo {author} {\bibfnamefont {A.~A.}\ \bibnamefont
  {Kovalev}}, \bibinfo {author} {\bibfnamefont {A.}~\bibnamefont {De}}, \ and\
  \bibinfo {author} {\bibfnamefont {K.}~\bibnamefont {Shtengel}},\ }\href
  {\doibase 10.1103/PhysRevLett.112.106402} {\bibfield  {journal} {\bibinfo
  {journal} {Phys. Rev. Lett.}\ }\textbf {\bibinfo {volume} {112}},\ \bibinfo
  {pages} {106402} (\bibinfo {year} {2014})}\BibitemShut {NoStop}%
\bibitem [{\citenamefont {Soykal}\ \emph {et~al.}(2011)\citenamefont {Soykal},
  \citenamefont {Ruskov},\ and\ \citenamefont {Tahan}}]{Soykal2011}%
  \BibitemOpen
  \bibfield  {author} {\bibinfo {author} {\bibfnamefont {O.~O.}\ \bibnamefont
  {Soykal}}, \bibinfo {author} {\bibfnamefont {R.}~\bibnamefont {Ruskov}}, \
  and\ \bibinfo {author} {\bibfnamefont {C.}~\bibnamefont {Tahan}},\ }\href
  {\doibase 10.1103/PhysRevLett.107.235502} {\bibfield  {journal} {\bibinfo
  {journal} {Phys. Rev. Lett.}\ }\textbf {\bibinfo {volume} {107}},\ \bibinfo
  {pages} {235502} (\bibinfo {year} {2011})}\BibitemShut {NoStop}%
\bibitem [{\citenamefont {Devoret}\ \emph {et~al.}(1989)\citenamefont
  {Devoret}, \citenamefont {Esteve}, \citenamefont {Martinis},\ and\
  \citenamefont {Urbina}}]{Devoret1989}%
  \BibitemOpen
  \bibfield  {author} {\bibinfo {author} {\bibfnamefont {M.~H.}\ \bibnamefont
  {Devoret}}, \bibinfo {author} {\bibfnamefont {D.}~\bibnamefont {Esteve}},
  \bibinfo {author} {\bibfnamefont {J.~M.}\ \bibnamefont {Martinis}}, \ and\
  \bibinfo {author} {\bibfnamefont {C.}~\bibnamefont {Urbina}},\ }\href@noop {}
  {\bibfield  {journal} {\bibinfo  {journal} {Physica Scripta}\ }\textbf
  {\bibinfo {volume} {1989}},\ \bibinfo {pages} {118} (\bibinfo {year}
  {1989})}\BibitemShut {NoStop}%
\bibitem [{\citenamefont {Schoelkopf}\ and\ \citenamefont
  {Girvin}(2008)}]{Schoelkopf2008}%
  \BibitemOpen
  \bibfield  {author} {\bibinfo {author} {\bibfnamefont {R.~J.}\ \bibnamefont
  {Schoelkopf}}\ and\ \bibinfo {author} {\bibfnamefont {S.~M.}\ \bibnamefont
  {Girvin}},\ }\href {\doibase 10.1038/451664a} {\bibfield  {journal} {\bibinfo
   {journal} {Nature}\ }\textbf {\bibinfo {volume} {451}},\ \bibinfo {pages}
  {664} (\bibinfo {year} {2008})}\BibitemShut {NoStop}%
\bibitem [{\citenamefont {Martinis}(2009)}]{Martinis2009}%
  \BibitemOpen
  \bibfield  {author} {\bibinfo {author} {\bibfnamefont {J.~M.}\ \bibnamefont
  {Martinis}},\ }\href {\doibase 10.1007/s11128-009-0105-1} {\bibfield
  {journal} {\bibinfo  {journal} {Quantum Information Processing}\ }\textbf
  {\bibinfo {volume} {8}},\ \bibinfo {pages} {81} (\bibinfo {year}
  {2009})}\BibitemShut {NoStop}%
\bibitem [{\citenamefont {Jones}\ \emph {et~al.}(2013)\citenamefont {Jones},
  \citenamefont {Yu}, \citenamefont {Ghimire}, \citenamefont {Wu},
  \citenamefont {Aivazian}, \citenamefont {Ross}, \citenamefont {Zhao},
  \citenamefont {Yan}, \citenamefont {Mandrus}, \citenamefont {Xiao} \emph
  {et~al.}}]{Jones2013}%
  \BibitemOpen
  \bibfield  {author} {\bibinfo {author} {\bibfnamefont {A.~M.}\ \bibnamefont
  {Jones}}, \bibinfo {author} {\bibfnamefont {H.}~\bibnamefont {Yu}}, \bibinfo
  {author} {\bibfnamefont {N.~J.}\ \bibnamefont {Ghimire}}, \bibinfo {author}
  {\bibfnamefont {S.}~\bibnamefont {Wu}}, \bibinfo {author} {\bibfnamefont
  {G.}~\bibnamefont {Aivazian}}, \bibinfo {author} {\bibfnamefont {J.~S.}\
  \bibnamefont {Ross}}, \bibinfo {author} {\bibfnamefont {B.}~\bibnamefont
  {Zhao}}, \bibinfo {author} {\bibfnamefont {J.}~\bibnamefont {Yan}}, \bibinfo
  {author} {\bibfnamefont {D.~G.}\ \bibnamefont {Mandrus}}, \bibinfo {author}
  {\bibfnamefont {D.}~\bibnamefont {Xiao}},  \emph {et~al.},\ }\href@noop {}
  {\bibfield  {journal} {\bibinfo  {journal} {Nature nanotechnology}\ }\textbf
  {\bibinfo {volume} {8}},\ \bibinfo {pages} {634} (\bibinfo {year}
  {2013})}\BibitemShut {NoStop}%
\bibitem [{\citenamefont {Xu}\ \emph {et~al.}(2014)\citenamefont {Xu},
  \citenamefont {Yao}, \citenamefont {Xiao},\ and\ \citenamefont
  {Heinz}}]{Xu2014}%
  \BibitemOpen
  \bibfield  {author} {\bibinfo {author} {\bibfnamefont {X.}~\bibnamefont
  {Xu}}, \bibinfo {author} {\bibfnamefont {W.}~\bibnamefont {Yao}}, \bibinfo
  {author} {\bibfnamefont {D.}~\bibnamefont {Xiao}}, \ and\ \bibinfo {author}
  {\bibfnamefont {T.~F.}\ \bibnamefont {Heinz}},\ }\href@noop {} {\bibfield
  {journal} {\bibinfo  {journal} {Nature Physics}\ }\textbf {\bibinfo {volume}
  {10}},\ \bibinfo {pages} {343} (\bibinfo {year} {2014})}\BibitemShut
  {NoStop}%
\bibitem [{\citenamefont {Jones}\ \emph {et~al.}(2014)\citenamefont {Jones},
  \citenamefont {Yu}, \citenamefont {Ross}, \citenamefont {Klement},
  \citenamefont {Ghimire}, \citenamefont {Yan}, \citenamefont {Mandrus},
  \citenamefont {Yao},\ and\ \citenamefont {Xu}}]{Jones2014}%
  \BibitemOpen
  \bibfield  {author} {\bibinfo {author} {\bibfnamefont {A.~M.}\ \bibnamefont
  {Jones}}, \bibinfo {author} {\bibfnamefont {H.}~\bibnamefont {Yu}}, \bibinfo
  {author} {\bibfnamefont {J.~S.}\ \bibnamefont {Ross}}, \bibinfo {author}
  {\bibfnamefont {P.}~\bibnamefont {Klement}}, \bibinfo {author} {\bibfnamefont
  {N.~J.}\ \bibnamefont {Ghimire}}, \bibinfo {author} {\bibfnamefont
  {J.}~\bibnamefont {Yan}}, \bibinfo {author} {\bibfnamefont {D.~G.}\
  \bibnamefont {Mandrus}}, \bibinfo {author} {\bibfnamefont {W.}~\bibnamefont
  {Yao}}, \ and\ \bibinfo {author} {\bibfnamefont {X.}~\bibnamefont {Xu}},\
  }\href@noop {} {\bibfield  {journal} {\bibinfo  {journal} {Nature Physics}\
  }\textbf {\bibinfo {volume} {10}},\ \bibinfo {pages} {130} (\bibinfo {year}
  {2014})}\BibitemShut {NoStop}%
\bibitem [{\citenamefont {He}\ \emph {et~al.}(2014)\citenamefont {He},
  \citenamefont {Kumar}, \citenamefont {Zhao}, \citenamefont {Wang},
  \citenamefont {Mak}, \citenamefont {Zhao},\ and\ \citenamefont
  {Shan}}]{He2014PRL}%
  \BibitemOpen
  \bibfield  {author} {\bibinfo {author} {\bibfnamefont {K.}~\bibnamefont
  {He}}, \bibinfo {author} {\bibfnamefont {N.}~\bibnamefont {Kumar}}, \bibinfo
  {author} {\bibfnamefont {L.}~\bibnamefont {Zhao}}, \bibinfo {author}
  {\bibfnamefont {Z.}~\bibnamefont {Wang}}, \bibinfo {author} {\bibfnamefont
  {K.~F.}\ \bibnamefont {Mak}}, \bibinfo {author} {\bibfnamefont
  {H.}~\bibnamefont {Zhao}}, \ and\ \bibinfo {author} {\bibfnamefont
  {J.}~\bibnamefont {Shan}},\ }\href {\doibase 10.1103/PhysRevLett.113.026803}
  {\bibfield  {journal} {\bibinfo  {journal} {Phys. Rev. Lett.}\ }\textbf
  {\bibinfo {volume} {113}},\ \bibinfo {pages} {026803} (\bibinfo {year}
  {2014})}\BibitemShut {NoStop}%
\bibitem [{\citenamefont {Chernikov}\ \emph {et~al.}(2014)\citenamefont
  {Chernikov}, \citenamefont {Berkelbach}, \citenamefont {Hill}, \citenamefont
  {Rigosi}, \citenamefont {Li}, \citenamefont {Aslan}, \citenamefont
  {Reichman}, \citenamefont {Hybertsen},\ and\ \citenamefont
  {Heinz}}]{Chernikov2014PRL}%
  \BibitemOpen
  \bibfield  {author} {\bibinfo {author} {\bibfnamefont {A.}~\bibnamefont
  {Chernikov}}, \bibinfo {author} {\bibfnamefont {T.~C.}\ \bibnamefont
  {Berkelbach}}, \bibinfo {author} {\bibfnamefont {H.~M.}\ \bibnamefont
  {Hill}}, \bibinfo {author} {\bibfnamefont {A.}~\bibnamefont {Rigosi}},
  \bibinfo {author} {\bibfnamefont {Y.}~\bibnamefont {Li}}, \bibinfo {author}
  {\bibfnamefont {O.~B.}\ \bibnamefont {Aslan}}, \bibinfo {author}
  {\bibfnamefont {D.~R.}\ \bibnamefont {Reichman}}, \bibinfo {author}
  {\bibfnamefont {M.~S.}\ \bibnamefont {Hybertsen}}, \ and\ \bibinfo {author}
  {\bibfnamefont {T.~F.}\ \bibnamefont {Heinz}},\ }\href {\doibase
  10.1103/PhysRevLett.113.076802} {\bibfield  {journal} {\bibinfo  {journal}
  {Phys. Rev. Lett.}\ }\textbf {\bibinfo {volume} {113}},\ \bibinfo {pages}
  {076802} (\bibinfo {year} {2014})}\BibitemShut {NoStop}%
\bibitem [{\citenamefont {Korm\'anyos}\ \emph {et~al.}(2014)\citenamefont
  {Korm\'anyos}, \citenamefont {Z\'olyomi}, \citenamefont {Drummond},\ and\
  \citenamefont {Burkard}}]{Kormanyos2014}%
  \BibitemOpen
  \bibfield  {author} {\bibinfo {author} {\bibfnamefont {A.}~\bibnamefont
  {Korm\'anyos}}, \bibinfo {author} {\bibfnamefont {V.}~\bibnamefont
  {Z\'olyomi}}, \bibinfo {author} {\bibfnamefont {N.~D.}\ \bibnamefont
  {Drummond}}, \ and\ \bibinfo {author} {\bibfnamefont {G.}~\bibnamefont
  {Burkard}},\ }\href {\doibase 10.1103/PhysRevX.4.011034} {\bibfield
  {journal} {\bibinfo  {journal} {Phys. Rev. X}\ }\textbf {\bibinfo {volume}
  {4}},\ \bibinfo {pages} {011034} (\bibinfo {year} {2014})}\BibitemShut
  {NoStop}%
\bibitem [{\citenamefont {Gong}\ \emph {et~al.}(2013)\citenamefont {Gong},
  \citenamefont {Liu}, \citenamefont {Yu}, \citenamefont {Xiao}, \citenamefont
  {Cui}, \citenamefont {Xu},\ and\ \citenamefont {Yao}}]{Gong2013}%
  \BibitemOpen
  \bibfield  {author} {\bibinfo {author} {\bibfnamefont {Z.}~\bibnamefont
  {Gong}}, \bibinfo {author} {\bibfnamefont {G.-B.}\ \bibnamefont {Liu}},
  \bibinfo {author} {\bibfnamefont {H.}~\bibnamefont {Yu}}, \bibinfo {author}
  {\bibfnamefont {D.}~\bibnamefont {Xiao}}, \bibinfo {author} {\bibfnamefont
  {X.}~\bibnamefont {Cui}}, \bibinfo {author} {\bibfnamefont {X.}~\bibnamefont
  {Xu}}, \ and\ \bibinfo {author} {\bibfnamefont {W.}~\bibnamefont {Yao}},\
  }\href@noop {} {\bibfield  {journal} {\bibinfo  {journal} {Nature
  communications}\ }\textbf {\bibinfo {volume} {4}},\ \bibinfo {pages} {2053}
  (\bibinfo {year} {2013})}\BibitemShut {NoStop}%
\bibitem [{\citenamefont {Gan}\ \emph {et~al.}(2012)\citenamefont {Gan},
  \citenamefont {Mak}, \citenamefont {Gao}, \citenamefont {You}, \citenamefont
  {Hatami}, \citenamefont {Hone}, \citenamefont {Heinz},\ and\ \citenamefont
  {Englund}}]{Gan2012}%
  \BibitemOpen
  \bibfield  {author} {\bibinfo {author} {\bibfnamefont {X.}~\bibnamefont
  {Gan}}, \bibinfo {author} {\bibfnamefont {K.~F.}\ \bibnamefont {Mak}},
  \bibinfo {author} {\bibfnamefont {Y.}~\bibnamefont {Gao}}, \bibinfo {author}
  {\bibfnamefont {Y.}~\bibnamefont {You}}, \bibinfo {author} {\bibfnamefont
  {F.}~\bibnamefont {Hatami}}, \bibinfo {author} {\bibfnamefont
  {J.}~\bibnamefont {Hone}}, \bibinfo {author} {\bibfnamefont {T.~F.}\
  \bibnamefont {Heinz}}, \ and\ \bibinfo {author} {\bibfnamefont
  {D.}~\bibnamefont {Englund}},\ }\href@noop {} {\bibfield  {journal} {\bibinfo
   {journal} {Nano letters}\ }\textbf {\bibinfo {volume} {12}},\ \bibinfo
  {pages} {5626} (\bibinfo {year} {2012})}\BibitemShut {NoStop}%
\bibitem [{\citenamefont {Koppens}\ \emph {et~al.}(2011)\citenamefont
  {Koppens}, \citenamefont {Chang},\ and\ \citenamefont {Garcia~de
  Abajo}}]{Koppens2011}%
  \BibitemOpen
  \bibfield  {author} {\bibinfo {author} {\bibfnamefont {F.~H.}\ \bibnamefont
  {Koppens}}, \bibinfo {author} {\bibfnamefont {D.~E.}\ \bibnamefont {Chang}},
  \ and\ \bibinfo {author} {\bibfnamefont {F.~J.}\ \bibnamefont {Garcia~de
  Abajo}},\ }\href@noop {} {\bibfield  {journal} {\bibinfo  {journal} {Nano
  letters}\ }\textbf {\bibinfo {volume} {11}},\ \bibinfo {pages} {3370}
  (\bibinfo {year} {2011})}\BibitemShut {NoStop}%
\bibitem [{\citenamefont {Britnell}\ \emph {et~al.}(2013)\citenamefont
  {Britnell}, \citenamefont {Ribeiro}, \citenamefont {Eckmann}, \citenamefont
  {Jalil}, \citenamefont {Belle}, \citenamefont {Mishchenko}, \citenamefont
  {Kim}, \citenamefont {Gorbachev}, \citenamefont {Georgiou}, \citenamefont
  {Morozov} \emph {et~al.}}]{Britnell2013}%
  \BibitemOpen
  \bibfield  {author} {\bibinfo {author} {\bibfnamefont {L.}~\bibnamefont
  {Britnell}}, \bibinfo {author} {\bibfnamefont {R.}~\bibnamefont {Ribeiro}},
  \bibinfo {author} {\bibfnamefont {A.}~\bibnamefont {Eckmann}}, \bibinfo
  {author} {\bibfnamefont {R.}~\bibnamefont {Jalil}}, \bibinfo {author}
  {\bibfnamefont {B.}~\bibnamefont {Belle}}, \bibinfo {author} {\bibfnamefont
  {A.}~\bibnamefont {Mishchenko}}, \bibinfo {author} {\bibfnamefont {Y.-J.}\
  \bibnamefont {Kim}}, \bibinfo {author} {\bibfnamefont {R.}~\bibnamefont
  {Gorbachev}}, \bibinfo {author} {\bibfnamefont {T.}~\bibnamefont {Georgiou}},
  \bibinfo {author} {\bibfnamefont {S.}~\bibnamefont {Morozov}},  \emph
  {et~al.},\ }\href@noop {} {\bibfield  {journal} {\bibinfo  {journal}
  {Science}\ }\textbf {\bibinfo {volume} {340}},\ \bibinfo {pages} {1311}
  (\bibinfo {year} {2013})}\BibitemShut {NoStop}%
\bibitem [{\citenamefont {Ye}\ \emph {et~al.}(2015)\citenamefont {Ye},
  \citenamefont {Wong}, \citenamefont {Lu}, \citenamefont {Zhu}, \citenamefont
  {Chen}, \citenamefont {Wang},\ and\ \citenamefont {Zhang}}]{Ye2015Nature}%
  \BibitemOpen
  \bibfield  {author} {\bibinfo {author} {\bibfnamefont {Y.}~\bibnamefont
  {Ye}}, \bibinfo {author} {\bibfnamefont {Z.~J.}\ \bibnamefont {Wong}},
  \bibinfo {author} {\bibfnamefont {X.}~\bibnamefont {Lu}}, \bibinfo {author}
  {\bibfnamefont {H.}~\bibnamefont {Zhu}}, \bibinfo {author} {\bibfnamefont
  {X.}~\bibnamefont {Chen}}, \bibinfo {author} {\bibfnamefont {Y.}~\bibnamefont
  {Wang}}, \ and\ \bibinfo {author} {\bibfnamefont {X.}~\bibnamefont {Zhang}},\
  }\href@noop {} {\bibfield  {journal} {\bibinfo  {journal} {Nature Photonics}\
  ,\ \bibinfo {pages} {733–737}} (\bibinfo {year} {2015})}\BibitemShut
  {NoStop}%
\bibitem [{\citenamefont {Xiao}\ \emph {et~al.}(2012)\citenamefont {Xiao},
  \citenamefont {Liu}, \citenamefont {Feng}, \citenamefont {Xu},\ and\
  \citenamefont {Yao}}]{Xiao2012PRL}%
  \BibitemOpen
  \bibfield  {author} {\bibinfo {author} {\bibfnamefont {D.}~\bibnamefont
  {Xiao}}, \bibinfo {author} {\bibfnamefont {G.-B.}\ \bibnamefont {Liu}},
  \bibinfo {author} {\bibfnamefont {W.}~\bibnamefont {Feng}}, \bibinfo {author}
  {\bibfnamefont {X.}~\bibnamefont {Xu}}, \ and\ \bibinfo {author}
  {\bibfnamefont {W.}~\bibnamefont {Yao}},\ }\href {\doibase
  10.1103/PhysRevLett.108.196802} {\bibfield  {journal} {\bibinfo  {journal}
  {Phys. Rev. Lett.}\ }\textbf {\bibinfo {volume} {108}},\ \bibinfo {pages}
  {196802} (\bibinfo {year} {2012})}\BibitemShut {NoStop}%
\bibitem [{\citenamefont {Hodgkinson}\ \emph {et~al.}(2002)\citenamefont
  {Hodgkinson}, \citenamefont {Arnold}, \citenamefont {McCall}, \citenamefont
  {Lakhtakia} \emph {et~al.}}]{Hodgkinson2002}%
  \BibitemOpen
  \bibfield  {author} {\bibinfo {author} {\bibfnamefont {I.~J.}\ \bibnamefont
  {Hodgkinson}}, \bibinfo {author} {\bibfnamefont {M.}~\bibnamefont {Arnold}},
  \bibinfo {author} {\bibfnamefont {M.~W.}\ \bibnamefont {McCall}}, \bibinfo
  {author} {\bibfnamefont {A.}~\bibnamefont {Lakhtakia}},  \emph {et~al.},\
  }\href@noop {} {\bibfield  {journal} {\bibinfo  {journal} {Optics
  communications}\ }\textbf {\bibinfo {volume} {210}},\ \bibinfo {pages} {201}
  (\bibinfo {year} {2002})}\BibitemShut {NoStop}%
\bibitem [{\citenamefont {Maksimov}\ \emph {et~al.}(2014)\citenamefont
  {Maksimov}, \citenamefont {Tartakovskii}, \citenamefont {Filatov},
  \citenamefont {Lobanov}, \citenamefont {Gippius}, \citenamefont {Tikhodeev},
  \citenamefont {Schneider}, \citenamefont {Kamp}, \citenamefont {Maier},
  \citenamefont {H\"ofling},\ and\ \citenamefont {Kulakovskii}}]{Maksimov2014}%
  \BibitemOpen
  \bibfield  {author} {\bibinfo {author} {\bibfnamefont {A.~A.}\ \bibnamefont
  {Maksimov}}, \bibinfo {author} {\bibfnamefont {I.~I.}\ \bibnamefont
  {Tartakovskii}}, \bibinfo {author} {\bibfnamefont {E.~V.}\ \bibnamefont
  {Filatov}}, \bibinfo {author} {\bibfnamefont {S.~V.}\ \bibnamefont
  {Lobanov}}, \bibinfo {author} {\bibfnamefont {N.~A.}\ \bibnamefont
  {Gippius}}, \bibinfo {author} {\bibfnamefont {S.~G.}\ \bibnamefont
  {Tikhodeev}}, \bibinfo {author} {\bibfnamefont {C.}~\bibnamefont
  {Schneider}}, \bibinfo {author} {\bibfnamefont {M.}~\bibnamefont {Kamp}},
  \bibinfo {author} {\bibfnamefont {S.}~\bibnamefont {Maier}}, \bibinfo
  {author} {\bibfnamefont {S.}~\bibnamefont {H\"ofling}}, \ and\ \bibinfo
  {author} {\bibfnamefont {V.~D.}\ \bibnamefont {Kulakovskii}},\ }\href
  {\doibase 10.1103/PhysRevB.89.045316} {\bibfield  {journal} {\bibinfo
  {journal} {Phys. Rev. B}\ }\textbf {\bibinfo {volume} {89}},\ \bibinfo
  {pages} {045316} (\bibinfo {year} {2014})}\BibitemShut {NoStop}%
\bibitem [{\citenamefont {Konishi}\ \emph {et~al.}(2011)\citenamefont
  {Konishi}, \citenamefont {Nomura}, \citenamefont {Kumagai}, \citenamefont
  {Iwamoto}, \citenamefont {Arakawa},\ and\ \citenamefont
  {Kuwata-Gonokami}}]{Konishi2011}%
  \BibitemOpen
  \bibfield  {author} {\bibinfo {author} {\bibfnamefont {K.}~\bibnamefont
  {Konishi}}, \bibinfo {author} {\bibfnamefont {M.}~\bibnamefont {Nomura}},
  \bibinfo {author} {\bibfnamefont {N.}~\bibnamefont {Kumagai}}, \bibinfo
  {author} {\bibfnamefont {S.}~\bibnamefont {Iwamoto}}, \bibinfo {author}
  {\bibfnamefont {Y.}~\bibnamefont {Arakawa}}, \ and\ \bibinfo {author}
  {\bibfnamefont {M.}~\bibnamefont {Kuwata-Gonokami}},\ }\href {\doibase
  10.1103/PhysRevLett.106.057402} {\bibfield  {journal} {\bibinfo  {journal}
  {Phys. Rev. Lett.}\ }\textbf {\bibinfo {volume} {106}},\ \bibinfo {pages}
  {057402} (\bibinfo {year} {2011})}\BibitemShut {NoStop}%
\bibitem [{\citenamefont {Wang}\ \emph {et~al.}(2015)\citenamefont {Wang},
  \citenamefont {Bouet}, \citenamefont {Glazov}, \citenamefont {Amand},
  \citenamefont {Ivchenko}, \citenamefont {Palleau}, \citenamefont {Marie},\
  and\ \citenamefont {Urbaszek}}]{Wang20152D}%
  \BibitemOpen
  \bibfield  {author} {\bibinfo {author} {\bibfnamefont {G.}~\bibnamefont
  {Wang}}, \bibinfo {author} {\bibfnamefont {L.}~\bibnamefont {Bouet}},
  \bibinfo {author} {\bibfnamefont {M.~M.}\ \bibnamefont {Glazov}}, \bibinfo
  {author} {\bibfnamefont {T.}~\bibnamefont {Amand}}, \bibinfo {author}
  {\bibfnamefont {E.~L.}\ \bibnamefont {Ivchenko}}, \bibinfo {author}
  {\bibfnamefont {E.}~\bibnamefont {Palleau}}, \bibinfo {author} {\bibfnamefont
  {X.}~\bibnamefont {Marie}}, \ and\ \bibinfo {author} {\bibfnamefont
  {B.}~\bibnamefont {Urbaszek}},\ }\href@noop {} {\bibfield  {journal}
  {\bibinfo  {journal} {2D Materials}\ }\textbf {\bibinfo {volume} {2}},\
  \bibinfo {pages} {034002} (\bibinfo {year} {2015})}\BibitemShut {NoStop}%
\bibitem [{\citenamefont {Aivazian}\ \emph {et~al.}(2015)\citenamefont
  {Aivazian}, \citenamefont {Gong}, \citenamefont {Jones}, \citenamefont {Chu},
  \citenamefont {Yan}, \citenamefont {Mandrus}, \citenamefont {Zhang},
  \citenamefont {Cobden}, \citenamefont {Yao},\ and\ \citenamefont
  {Xu}}]{Aivazian2015Nat}%
  \BibitemOpen
  \bibfield  {author} {\bibinfo {author} {\bibfnamefont {G.}~\bibnamefont
  {Aivazian}}, \bibinfo {author} {\bibfnamefont {Z.}~\bibnamefont {Gong}},
  \bibinfo {author} {\bibfnamefont {A.~M.}\ \bibnamefont {Jones}}, \bibinfo
  {author} {\bibfnamefont {R.-L.}\ \bibnamefont {Chu}}, \bibinfo {author}
  {\bibfnamefont {J.}~\bibnamefont {Yan}}, \bibinfo {author} {\bibfnamefont
  {D.~G.}\ \bibnamefont {Mandrus}}, \bibinfo {author} {\bibfnamefont
  {C.}~\bibnamefont {Zhang}}, \bibinfo {author} {\bibfnamefont
  {D.}~\bibnamefont {Cobden}}, \bibinfo {author} {\bibfnamefont
  {W.}~\bibnamefont {Yao}}, \ and\ \bibinfo {author} {\bibfnamefont
  {X.}~\bibnamefont {Xu}},\ }\href@noop {} {\bibfield  {journal} {\bibinfo
  {journal} {Nature Physics}\ ,\ \bibinfo {pages} {148–152}} (\bibinfo {year}
  {2015})}\BibitemShut {NoStop}%
\bibitem [{\citenamefont {Zibouche}\ \emph {et~al.}(2014)\citenamefont
  {Zibouche}, \citenamefont {Kuc}, \citenamefont {Musfeldt},\ and\
  \citenamefont {Heine}}]{Heine_AnnPhys14}%
  \BibitemOpen
  \bibfield  {author} {\bibinfo {author} {\bibfnamefont {N.}~\bibnamefont
  {Zibouche}}, \bibinfo {author} {\bibfnamefont {A.}~\bibnamefont {Kuc}},
  \bibinfo {author} {\bibfnamefont {J.}~\bibnamefont {Musfeldt}}, \ and\
  \bibinfo {author} {\bibfnamefont {T.}~\bibnamefont {Heine}},\ }\href
  {\doibase 10.1002/andp.201400137} {\bibfield  {journal} {\bibinfo  {journal}
  {Ann. Phys. (Berlin)}\ }\textbf {\bibinfo {volume} {526}},\ \bibinfo {pages}
  {395 } (\bibinfo {year} {2014})}\BibitemShut {NoStop}%
\bibitem [{\citenamefont {Korm{\'a}nyos}\ \emph {et~al.}(2015)\citenamefont
  {Korm{\'a}nyos}, \citenamefont {Burkard}, \citenamefont {Gmitra},
  \citenamefont {Fabian}, \citenamefont {Z{\'o}lyomi}, \citenamefont
  {Drummond},\ and\ \citenamefont {Fal’ko}}]{Kormanyos2015}%
  \BibitemOpen
  \bibfield  {author} {\bibinfo {author} {\bibfnamefont {A.}~\bibnamefont
  {Korm{\'a}nyos}}, \bibinfo {author} {\bibfnamefont {G.}~\bibnamefont
  {Burkard}}, \bibinfo {author} {\bibfnamefont {M.}~\bibnamefont {Gmitra}},
  \bibinfo {author} {\bibfnamefont {J.}~\bibnamefont {Fabian}}, \bibinfo
  {author} {\bibfnamefont {V.}~\bibnamefont {Z{\'o}lyomi}}, \bibinfo {author}
  {\bibfnamefont {N.~D.}\ \bibnamefont {Drummond}}, \ and\ \bibinfo {author}
  {\bibfnamefont {V.}~\bibnamefont {Fal’ko}},\ }\href@noop {} {\bibfield
  {journal} {\bibinfo  {journal} {2D Materials}\ }\textbf {\bibinfo {volume}
  {2}},\ \bibinfo {pages} {022001} (\bibinfo {year} {2015})}\BibitemShut
  {NoStop}%
\end{thebibliography}
%

\end{document}